\documentclass[final,english]{bullsrsl}

\usepackage[latin1]{inputenc}

\usepackage{booktabs}
\usepackage{natbib} 
\usepackage{graphicx}

\begin{document}
\title{Optical observations of supernova candidates with the ILMT}

\author[affil={1,2}, corresponding]{Monalisa}{Dubey}
\author[affil={3,1}]{Bhavya}{Ailawadhi}
\author[affil={4,5}]{Talat}{Akhunov}
\author[affil={1,2}]{Naveen}{Dukiya}
\author[affil={6}]{Sara}{Filali}
\author[affil={7,8}]{Paul}{Hickson}
\author[affil={1}]{Priyanshi}{Kumari}
\author[affil={1}]{Gokul Singh}{Mehra}
\author[affil={1}]{Kuntal}{Misra}
\author[affil={9}]{Vibhore}{Negi}
\author[affil={1,10}]{Kumar}{Pranshu}
\author[affil={1,6}]{Jean}{Surdej}
\author[affil={1,10}]{Sarvesh Kumar}{Yadav}

\affiliation[1]{Aryabhatta Research Institute of observational sciencES (ARIES), Manora Peak, Nainital, 263001, India}
\affiliation[2]{Department of Applied Physics, Mahatma Jyotiba Phule Rohilkhand University, Bareilly, 243006, India}
\affiliation[3]{Physical Research Laboratory, Ahmedabad 380009, Gujarat, India}
\affiliation[4]{National University of Uzbekistan, Department of Astronomy and Astrophysics, 100174 Tashkent, Uzbekistan}
\affiliation[5]{Ulugh Beg Astronomical Institute of the Uzbek Academy of Sciences, Astronomicheskaya 33, 100052 Tashkent, Uzbekistan}
\affiliation[6]{Institute of Astrophysics and Geophysics, Li{\`e}ge University, All\'ee du 6 Ao\^ut 19c, 4000 Li{\`e}ge, Belgium}
\affiliation[7]{Department of Physics and Astronomy, The University of British Columbia, 6224 Agricultural Road, Vancouver, BC V6T 1Z1, Canada}
\affiliation[8]{Outer Space Institute, The University of British Columbia, 325-6224 Agricultural Road, Vancouver, BC V6T 1Z1, Canada}
\affiliation[9]{Kavli Institute for Astronomy and Astrophysics, Peking University, Beijing 100871, China
}
\affiliation[10]{Department of Applied Optics and Photonics, University of Calcutta, Kolkata, 700106, India}

\correspondance{monalisadubeyprl@gmail.com, monalisa@aries.res.in}

\maketitle

\begin{abstract}
The International Liquid Mirror Telescope (ILMT) is a 4-m optical ({\em gri} spectral bands) zenithal survey telescope with a field of view of 22.3$\times$22.3 arcmin, suitable for discovery and follow-up of faint transients. With the \texttt{PyLMT} pipeline, several new transients, including supernova candidates, are being discovered and detected routinely. We report the discovery of six supernova candidates, AT~2023yjc, AT~2024agkc, AT~2024aifv, AT~2024aiha, AT~2024fxn, and AT~2024zsm, using the ILMT. Ten additional supernova candidates were discovered in other surveys but were also detected with the ILMT. Here, we present the light curves of all supernova candidates observed with the ILMT in the {\em gri} bands. The basic light curve parameters, such as the rise time, decline rate, and peak magnitude, are estimated and reported. 
\end{abstract}

\keywords{Supernovae, Image subtraction, Photometry, Liquid Mirror Telescope}




\section{Introduction}
Supernovae (SNe) are one of the most violent explosions in the universe, arising either by the thermonuclear disruption of a White Dwarf \citep{Nomoto_1982}, or core collapse of a massive star \citep[$>$ 8 M$_\odot$;] []{Smartt_IIP_2009}. To understand the evolution of an SN and characterize its progenitor properties, systematic photometric and spectroscopic observations are required. However, for distant events, spectroscopic observations may not always be feasible; therefore, we must rely on photometric observations. The International Liquid Mirror Telescope (ILMT) is a 4-m survey telescope aimed to discover new SNe \citep{Surdej_2025} and monitor them with a daily cadence. It observes the zenith sky each night in a 22$'$ wide strip in SDSS {\em gri} bands. The limiting magnitude of 21.7 mag is reached in the {\em r} band; hence, faint transients can be captured by the telescope \citep{Ailawadhi_2024}. The telescope scans the full R.A. range from October to May, allowing us to obtain high-cadence observations \citep{Dubey_2024}. Over the past two years, ILMT has discovered six new SN candidates and observed ten additional SN candidates within its observable sky strip. For most candidates, spectroscopic observations were not possible due to their faint nature; hence, ILMT's one-day cadence light curves are a valuable tool for classifying and studying the evolution of these transients. 

In this work, we present the light curves of sixteen SN candidates observed with the ILMT and other surveys, including ZTF, ATLAS, and Pan-STARRS. The observations and data reduction procedure are discussed in Section~\ref{sec:obs}. The SN light curves and their parameters are presented in Section~\ref{sec: result}. Summary of the work is given in Section~\ref{sec: summary}.

\section{Observations and Data Reduction}
\label{sec:obs}
An automated image subtraction and convolutional neural network (CNN) based pipeline \citep[\texttt{PYLMT};][]{Pranshu_2025} has been integrated with the ILMT, which generates an alert upon the discovery of new transients. So far, ILMT has observed sixteen new SN candidates, among which six (AT~2023yjc; R.A = 01:50:02.90, Dec. = $+$29:08:53.00; \citealt{Pranshu_2023yjc}, AT~2024agkc; R.A = 07:49:08.10, Dec. = $+$29:35:07.00; \citealt{Pranshu_2024agkc}, AT~2024aifv; R.A = 16:12:11.10, Dec. = $+$29:34:17.00; \citealt{Pranshu_2024aifv}, AT~2024aiha; R.A = 05:39:53.40, Dec. = $+$29:29:04.00; \citealt{Pranshu_2024aiha}, AT~2024fxn; R.A = 14:13:49.60, Dec. = $+$29:23:32.00,$+$29:08:53.00; \citealt{Pranshu_2024fxn}, and AT~2024zsm; R.A = 02:52:40.00, Dec. = $+$29:10:09.00; \citealt{Pranshu_2024zsm}) are new discoveries. Because ILMT can detect fainter events, three SN candidates (AT~2024zsm, AT~2024agkc, and AT~2024aifv) were discovered exclusively with the ILMT.

\begin{figure}[t!]
    \centering  \includegraphics[width=\columnwidth]{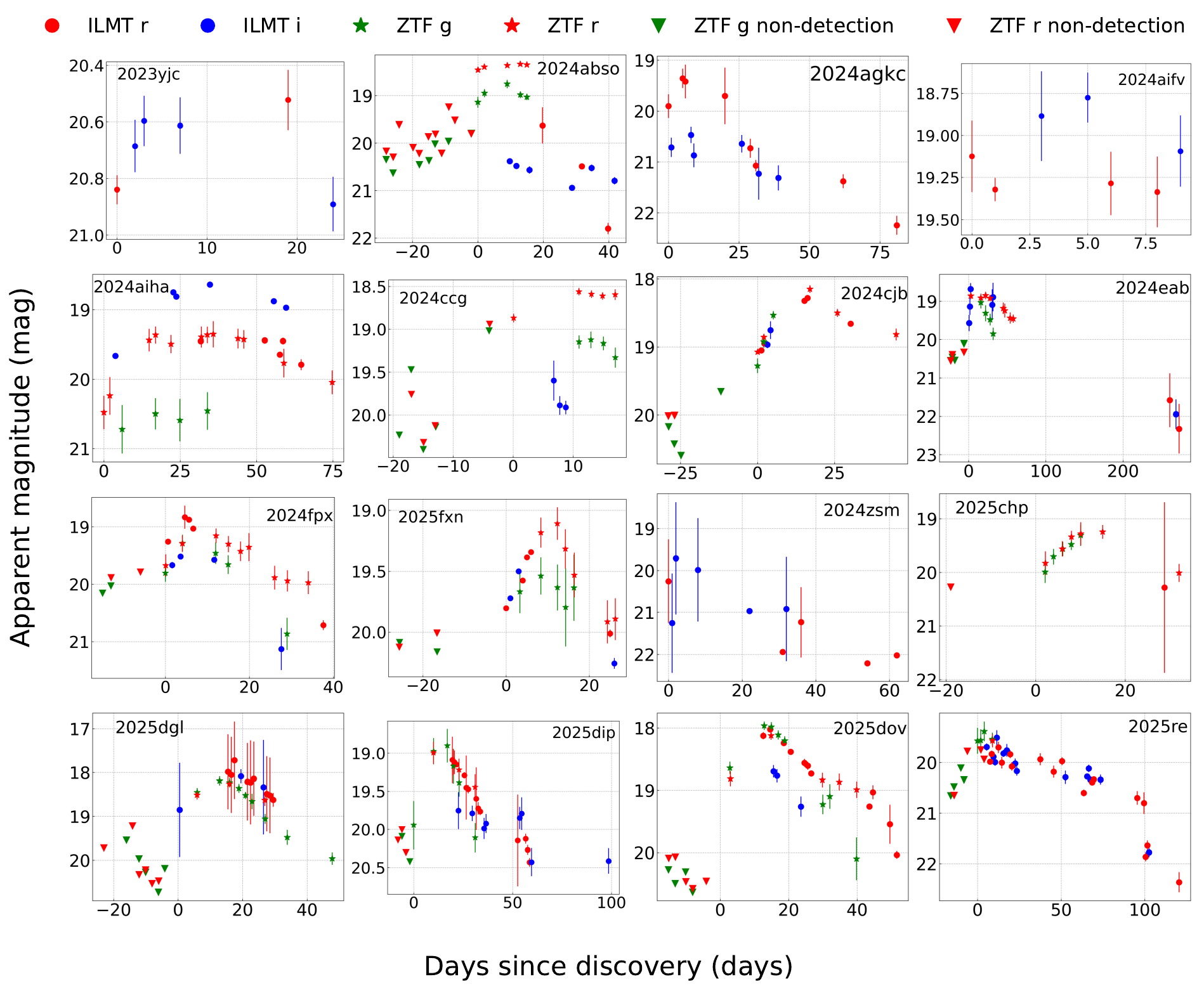}
    \begin{minipage}{12cm}
    \caption{The optical light curves of sixteen SN candidates constructed using the ILMT (circle) and publicly available ZTF data (asterisk) are shown. Non-detections in the ZTF are represented by green and red inverted triangles.}
    \label{fig:SN_lc}
    \end{minipage}
\end{figure}

The raw CCD frames acquired with the ILMT correspond to 4k $\times$ 36k pixels. In order to minimize the variation of the point spread function (PSF) along the declination axis, we selected a smaller sub-frame of 1k $\times$ 4k pixels centred on the SN candidate to perform the photometry. For the majority of the candidates, the host galaxy template images are taken approximately 1 year apart at the same Local Sidereal Time (LST) and have been used to perform image subtraction using the High Order Transform of PSF ANd Template Subtraction (HOTPANTS; \citep[\texttt{HOTPANTS};][]{Becker_2015}). Template image subtraction has not been performed for AT~2023yjc, AT~2024aiha, AT~2024cjb, AT~2024fpx, and AT~2025dov, as these are bright enough and are located at the outskirts of the host galaxy. PSF photometry and photometric calibration have been performed using a custom-built pipeline described in Dukiya et al. (in preparation). The photometric measurements of the SN candidates in {\em r} and {\em i} are available upon request. The ILMT light curves are supplemented with the ZTF {\em g} and {\em r} band data wherever available. The combined light curves are shown in \autoref{fig:SN_lc}.

\section{Results and Discussion}
\label{sec: result}

The light curves of all SN candidates are shown starting from the discovery time reported by the Transient Name Server (TNS). For AT~2025re, the data cover the longest observation duration (till $\sim$125 days). AT~2024cjb and AT~2025dov are spectroscopically classified as Type~Ia SN candidates by \cite{classification_2024cjb}, and \cite{classification_2025dov}, respectively. Apart from these two candidates, none of the others have spectroscopic classification; therefore, the analysis relies solely on photometric properties. The discovery details, detection information, along with the observational properties derived from the multi-band light curves, such as peak magnitude, rise time, and decline rates, are evaluated and summarized in \autoref{tab:sn_photo_prop}. The SN candidates are detected at a magnitude fainter than 18 mag in the {\em r} band. The observations cover the rise-to-peak in twelve candidates for the {\em r} and {\em i} bands. Due to insufficient data, the photometric parameters could not be estimated for some of the candidates.

The candidates exhibit a wide range of apparent peak magnitudes (17.9--20.2, 17.7--20.3, 18.1--20.6 mag), rise times ($\sim$ 4--35, $\sim$ 5--35, $\sim$2--35 days), and decline rates (0.013--0.094, 0.008--0.188, 0.004--0.079 mag day$^{-1}$) in the {\em gri} bands, highlighting the significant diversity in SN photometric properties. These spans are broadly consistent with the observed photometric properties of different well-studied SN types. For example, \cite{Papadogiannakis_2019} derived an average rise time in the {\em R} band of $\sim$16.8 days with an absolute peak brightness of $\sim$ $-$19 mag from a sample of 265 Type~Ia SNe. Among core-collapse SNe (CCSNe), Type IIP typically exhibit short rise times \citep[ $\leq$10 days;][]{Faran_2014} and shallow declines \citep[0.001--0.02 mag day$^{-1}$;][]{Anderson_2014, Sanders_2015} during the plateau phase, whereas interacting Type IIn SNe often show longer rise times \citep[$>$ 20 days;][]{Kiewe_2012} with steep decline \citep[0.013--0.04;][]{Nyholm_2020}. \cite{Taddia_2015} showed {\em r} band rise times vary for stripped envelope SNe (SESNe) subtypes, and typical values are $\sim$ 22.9 days for IIb, $\sim$ 21.3 days for Ib, $\sim$ 11.5 days for Ic, and $\sim$ 14.7 days for Ic broad-lined (Ic-BL) SNe. For some peculiar event like SN~1987A, the rise time was very long \citep[$>$ 40 days;][]{Taddia_2016}. The decline rate for SESNe ranges from 0.015 to 0.05 mag day$^{-1}$ \citep{Drout_2011}; however, some Type~Ic-BL SNe decline more rapidly, reflecting lower envelope mass \citep{Taddia_2016}. Since the light-curve parameters of our candidates fall within these established ranges for both Type~Ia and CCSNe, this supports their identification as SN candidates.

\begin{table*}[t]
\centering
\begin{minipage}{160mm}
\centering
\caption{Photometric properties of the SN candidates.}
\end{minipage}
\label{tab:sn_photo_prop}
\small

\resizebox{\textwidth}{!}{%
\begin{tabular}{l p{4.1cm} p{4.2cm} p{2.3cm} p{2.3cm} p{2.4cm}}
\toprule
SN &
Discovery date \& time (UT)$^\dagger$ &
Detection magnitude / No. of ILMT detections / Other sources &
Peak magnitude ($g/r/i$) &
Rise time ($g/r/i$) &
Decline rate ($g/r/i$) \\
\hline

AT 2023yjc &
2023-11-13T16:57:44 &
21.55 ($r$) / 6 / A &
\begin{tabular}[c]{@{}l@{}}--\\--\\20.60 (0.09)
\end{tabular} &
\begin{tabular}[c]{@{}l@{}}--\\--\\
3.00 (1.65)
\end{tabular} &
\begin{tabular}[c]{@{}l@{}}--\\--\\0.004 (0.033)
\end{tabular} \\
\hline

AT 2024abso &
2024-11-20T06:25:16.00 &
19.93 ($g$) / 9 / B, C &
\begin{tabular}[c]{@{}l@{}}19.24 (0.15)\\19.50 (0.14)\\20.38 (0.04)\end{tabular} &
\begin{tabular}[c]{@{}l@{}}8.95 (1.64)\\12.93 (7.10)\\9.74 (1.00)\end{tabular} &
\begin{tabular}[c]{@{}l@{}}0.094 (0.051)\\0.020 (0.091)\\0.050 (0.032)\end{tabular} \\
\hline

AT 2024agkc &
2024-12-30T19:49:40 &
19.46 ($r$) / 14 / None &
\begin{tabular}[c]{@{}l@{}}--\\19.36 (0.19)\\20.47 (0.16)\end{tabular} &
\begin{tabular}[c]{@{}l@{}}--\\5.00 (3.08)\\8.00 (0.64)\end{tabular} &
\begin{tabular}[c]{@{}l@{}}--\\0.023 (0.039)\\0.010 (0.013)\end{tabular} \\
\hline

AT 2024aifv &
2025-04-19T21:00:06 &
20.27 ($r$) / 7 / None &
\begin{tabular}[c]{@{}l@{}}--\\19.12 (0.21)\\18.78 (0.15)\end{tabular} &
\begin{tabular}[c]{@{}l@{}}--\\--\\5.00 (1.94)\end{tabular} &
\begin{tabular}[c]{@{}l@{}}--\\--\\0.079 (0.065)\end{tabular} \\
\hline

AT 2024aiha &
2024-10-10T23:02:31 &
20.37 ($r$) / 11 / C &
\begin{tabular}[c]{@{}l@{}}20.21 (0.13)\\19.35 (0.19)\\18.64 (0.02)\end{tabular} &
\begin{tabular}[c]{@{}l@{}}35.68 (0.95)\\35.80 (24.28)\\34.69 (2.20)\end{tabular} &
\begin{tabular}[c]{@{}l@{}}--\\0.008 (0.029)\\0.011 (0.002)\end{tabular} \\
\hline

AT 2024ccg &
2024-01-31T05:32:38.40 &
20.71 ($g$) / 4 / A &
\begin{tabular}[c]{@{}l@{}}19.59 (0.16)\\19.42 (0.13)\\19.60 (0.00)\end{tabular} &
\begin{tabular}[c]{@{}l@{}}12.96 (5.08)\\10.96 (4.28)\\6.77 (0.50)\end{tabular} &
\begin{tabular}[c]{@{}l@{}}0.031 (0.099)\\0.031 (0.091)\\--\end{tabular} \\
\hline

AT 2024cjb &
2024-02-12T06:52:09.00 &
20.08 ($g$) / 7 / A, B, C &
\begin{tabular}[c]{@{}l@{}}18.87 (0.09)\\18.28 (0.01)\\--\end{tabular} &
\begin{tabular}[c]{@{}l@{}}4.62 (0.01)\\16.22 (0.50)\\--\end{tabular} &
\begin{tabular}[c]{@{}l@{}}--\\0.188 (0.219)\\--\end{tabular} \\
\hline

AT 2024eab &
2024-03-06T09:56:28.61 &
19.42 ($w$-P1) / 8 / A, B, C &
\begin{tabular}[c]{@{}l@{}}19.04 (0.16)\\18.85 (0.10)\\18.68 (0.01)\end{tabular} &
\begin{tabular}[c]{@{}l@{}}15.78 (3.44)\\21.87 (8.28)\\2.59 (0.50)\end{tabular} &
\begin{tabular}[c]{@{}l@{}}0.046 (0.043)\\0.012 (0.021)\\0.015 (0.001)\end{tabular} \\
\hline

AT 2024fpx &
2024-04-04T09:10:23.998 &
19.8072 ($g$) / 9 / B, C &
\begin{tabular}[c]{@{}l@{}}19.29 (0.15)\\18.83 (0.01)\\19.52 (0.01)\end{tabular} &
\begin{tabular}[c]{@{}l@{}}4.09 (6.77)\\4.62 (0.50)\\3.62 (0.50)\end{tabular} &
\begin{tabular}[c]{@{}l@{}}0.022 (0.030)\\0.043 (0.001)\\--\end{tabular} \\
\hline

AT 2024fxn &
2024-04-05T00:00:00 &
19.74 ($r$) / 8 / A, B, C &
\begin{tabular}[c]{@{}l@{}}19.54 (0.15)\\19.11 (0.13)\\--\end{tabular} &
\begin{tabular}[c]{@{}l@{}}8.34 (6.49)\\12.30 (1.35)\\--\end{tabular} &
\begin{tabular}[c]{@{}l@{}}0.024 (0.061)\\0.057 (0.008)\\--\end{tabular} \\
\hline

AT 2024zsm &
2024-10-28T18:52:28 &
20.21 ($r$) / 10 / None &
\begin{tabular}[c]{@{}l@{}}--\\20.26 (1.00)\\19.71 (1.34)\end{tabular} &
\begin{tabular}[c]{@{}l@{}}--\\--\\2.00 (9.31)\end{tabular} &
\begin{tabular}[c]{@{}l@{}}--\\0.054 (0.032)\\0.046 (0.304)\end{tabular} \\
\hline

AT 2025chp &
2025-02-18T06:11:36.003 &
20.62 ($r$) / 1 / A, B, C, D &
\begin{tabular}[c]{@{}l@{}}19.31 (0.20)\\19.25 (0.13)\\--\end{tabular} &
\begin{tabular}[c]{@{}l@{}}10.04 (0.11)\\14.97 (1.78)\\--\end{tabular} &
\begin{tabular}[c]{@{}l@{}}--\\0.075 (0.009)\\--\end{tabular} \\
\hline

AT 2025dgl &
2025-03-03T12:49:09.69 &
19.05 ($o$) / 13 / B, C, E, F &
\begin{tabular}[c]{@{}l@{}}18.19 (0.11)\\17.72 (0.01)\\18.08 (0.01)\end{tabular} &
\begin{tabular}[c]{@{}l@{}}11.85 (8.29)\\16.47 (0.50)\\18.47 (0.50)\end{tabular} &
\begin{tabular}[c]{@{}l@{}}0.013 (0.047)\\0.068 (0.005)\\--\end{tabular} \\
\hline

AT 2025dip &
2025-02-27T09:39:04.99 &
19.95 ($g$) / 21 / A, B, C &
\begin{tabular}[c]{@{}l@{}}18.90 (0.22)\\19.00 (0.15)\\19.75 (0.01)\end{tabular} &
\begin{tabular}[c]{@{}l@{}}16.98 (2.61)\\--\\--\end{tabular} &
\begin{tabular}[c]{@{}l@{}}0.086 (0.102)\\0.010 (0.015)\\0.005 (0.001)\end{tabular} \\
\hline

AT 2025dov &
2025-03-06T10:11:19.39 &
19.29 ($o$) / 14 / A, B, C, E, F &
\begin{tabular}[c]{@{}l@{}}17.97 (0.07)\\18.03 (0.01)\\18.70 (0.01)\end{tabular} &
\begin{tabular}[c]{@{}l@{}}12.84 (7.72)\\14.58 (0.50)\\15.58 (0.50)\end{tabular} &
\begin{tabular}[c]{@{}l@{}}0.057 (0.027)\\0.055 (0.009)\\0.071 (0.146)\end{tabular} \\
\hline

AT 2025re &
2025-01-16T14:20:19.68 &
19.94 ($i$-P1) / 31 / A, C &
\begin{tabular}[c]{@{}l@{}}19.53 (0.22)\\19.70 (0.01)\\19.51 (0.01)\end{tabular} &
\begin{tabular}[c]{@{}l@{}}3.97 (6.18)\\12.47 (0.50)\\11.47 (0.50)\end{tabular} &
\begin{tabular}[c]{@{}l@{}}--\\0.009 (0.002)\\0.008 (0.002)\end{tabular} \\
\hline

\end{tabular}%
}

\footnotesize
\hspace{-5cm}A: Pan-STARRS; B: ATLAS; C: ZTF; D: BlackGEM; E: LAST; F: GOTO.\\
$^\dagger$ Discovery date and detection magnitude of each SN candidate are taken from the TNS, except for AT~2024aifv,\\\hspace{-1.2cm}whose discovery date is listed incorrectly in the TNS. The correct discovery date is given in this table.\par
\vspace{-2mm}
\rule{\linewidth}{\heavyrulewidth}

\end{table*}

\section{Summary}
\label{sec: summary}

In this work, we present multi-band optical light curves for sixteen SN candidates observed with the ILMT. Six of these are new discoveries exclusively with the ILMT. The basic photometric parameters, such as detection and peak magnitude, rise time, and decline rates, have been derived. Most of the detected SN candidates are fainter than 18 mag and could be detected thanks to the 4-m aperture of the ILMT. In the future, high-cadence photometric observations with the ILMT will allow the detection and monitoring of a larger number of SNe in distant galaxies.

\clearpage

\begin{acknowledgments}

The authors thank the referee for providing constructive comments on the manuscript, improving the clarity of presentation. We acknowledge the Belgo-Indian Network for Astronomy and astrophysics (BINA) consortium approved by the International Division, Department of Science and Technology (DST, Govt. of India; DST/INT/BELG/P-09/2017) and the Belgian Federal Science Policy Office (BELSPO, Govt. of Belgium; BL/33/IN12), for the local support to attend the fourth BINA workshop and allowing us to present our work in the form of a paper in the proceedings of the fourth BINA workshop. The 4-m International Liquid Mirror Telescope (ILMT) project results from a collaboration between the Institute of Astrophysics and Geophysics (University of Li\`{e}ge, Belgium), the Universities of British Columbia, Laval, Montreal, Toronto, Victoria, and York University, and Aryabhatta Research Institute of Observational Sciences (ARIES, India). The team acknowledges the contributions of ARIES's technical and scientific staff. JS wishes to thank Service Public Wallonie, F.R.S.-FNRS (Belgium), and the University of Li\`{e}ge, Belgium, for funding the construction of the ILMT. PH acknowledges financial support from the Natural Sciences and Engineering Research Council of Canada, RGPIN-2019-04369. PH, SF, and JS thank ARIES for hospitality during their visits to Devasthal. We sincerely acknowledge Anna Pospiezalska for her significant contributions to ILMT's functioning. MD acknowledges the Innovation in Science Pursuit for Inspired Research (INSPIRE) fellowship award (DST/INSPIRE Fellowship/2020/IF200251) for this work. KP acknowledges the support from the Erasmus+ Programme of the European Union for a research visit to the Institute of Astrophysics and Geophysics, University of Li\`ege, Belgium. TA thanks the Ministry of Higher Education, Science, and Innovations of Uzbekistan (grant FZ-20200929344). JS and KM acknowledge the assistance received from the ANRF SERB-726 VAJRA Faculty Scheme, India. KM acknowledges the support from the BRICS grant DST/ICD/BRICS/Call-5/CoNMuTraMO/2023 (G) funded by the DST, India.

\end{acknowledgments}

\begin{furtherinformation}

\begin{orcids}
\orcid{0009-0002-2621-6611}{Monalisa}{Dubey}
\orcid{0000-0001-5115-6310}{Talat} {Akhunov}
\orcid{0009-0000-1020-9711}{Bhavya} {Ailawadhi}
\orcid{0000-0002-0394-6745}{Naveen} {Dukiya}
\orcid{0000-0002-8289-8242}{Paul} {Hickson}
\orcid{0000-0003-1637-267X}{Kuntal}{Misra}
\orcid{0000-0001-5824-1040}{Vibhore} {Negi}
\orcid{0009-0005-3844-3426}{Kumar} {Pranshu}
\orcid{0000-0002-7005-1976}{Jean}{Surdej}

\end{orcids}

\begin{authorcontributions}
This work results from a long-term collaboration to which all authors have made significant contributions.

\end{authorcontributions}

\begin{conflictsofinterest}
The authors declare no conflict of interest.
\end{conflictsofinterest}

\end{furtherinformation}



%

\bibliographystyle{bullsrsl-en}

\bibliography{reference}

\end{document}